\documentclass{synmet}

\usepackage[dvips]{graphics}

\begin{document}

\def\mbi#1{\mbox{\boldmath$#1$}}
\def\kp{\mbi{k} \cdot \mbi{p}}
\def\eg{E_{\rm g}}
\def\eps{\varepsilon}
\def\la{\langle}
\def\ra{\rangle}
\def\beeq{\begin{equation}}
\def\eneq{\end{equation}}
\def\beeqa{\begin{eqnarray}}
\def\eneqa{\end{eqnarray}}
\def\theequation{\arabic{equation}}
\def\tmptheequation{\arabic{tmpequation}}
\renewcommand{\baselinestretch}{0.85}

\begin{frontmatter}



\title{Spin dependent transport of ``nonmagnetic metal/zigzag nanotube encapsulating magnetic atoms/nonmagnetic metal'' junctions}


\author{S. Kokado$^{a,b,}$
\thanksref{X}} and 
\author{K. Harigaya$^{b,c,}$ 
\thanksref{Y}}
\thanks[X]{Tel: +81-53-478-1644; Fax: +81-53-478-1644; 
E-mail: tskokad@ipc.shizuoka.ac.jp}
\thanks[Y]{Corresponding author. Tel: +81-29-861-5151; Fax: +81-29-861-5375; 
E-mail: k.harigaya@aist.go.jp}

\address{
$^a$Faculty of Engineering, Shizuoka University, 
Hamamatsu 432-8561, Japan\\
$^b$Nanotechnology Research Institute, AIST, Tsukuba 305-8568, Japan \\
$^c$Synthetic Nano-Function Materials Project, AIST, Tsukuba 305-8568, Japan
}

\begin{abstract}
Towards a novel magnetoresistance (MR) device with a carbon nanotube, 
we propose ``nonmagnetic metal/zigzag nanotube 
encapsulating magnetic atoms/nonmagnetic metal'' junctions. 
We theoretically investigate how spin-polarized edges of the nanotube 
and the encapsulated magnetic atoms influence on transport. 
When the on-site Coulomb energy divided 
by the magnitude of transfer integral, $U/|t|$, is larger than 0.8, 
large MR effect due to the direction of spins of magnetic atoms, 
which has the magnitude of the MR ratio of about 100\%, 
appears reflecting such spin-polarized edges. 
\end{abstract}
\end{frontmatter}
\renewcommand{\baselinestretch}{0.85}
%
%
%
%
\section{Introduction}
Spin dependent transport of junctions 
with a carbon nanotube~\cite{Tsukagoshi,Zhao,Zhao2,Kim,Mehrez,Kokado,kokado2} 
is one of the most interesting topics 
in nano-spintronics fields. 
In fact, since it was reported that 
ferromagnet (FM)/carbon nanotube/FM junctions 
exhibited the magnetoresistance (MR) effect~\cite{MR}
with the magnitude of the MR ratio of 9\% at 4.2 K~\cite{Tsukagoshi}, 
many experimental and theoretical studies 
on this system have been performed. 
For example, the recent experimental studies showed 
the magnitude of MR ratios of 
23\%~\cite{Zhao} and 26\%~\cite{Zhao2} at 4.2 K, 
and theoretical study 
based on the Green's function method showed that 
a maximum value of the MR ratio was evaluated as 20\%~\cite{Mehrez}. 
However, 
except for FM/carbon nanotube/FM 
junctions~\cite{Tsukagoshi,Zhao,Zhao2,Kim,Mehrez}, 
MR devices with the nanotube 
were hardly studied so far. 
If devices with larger MR ratio than the conventional ones can be 
successfully designed by exploiting characteristic in the nanotube, 
such the material design will significantly contribute to 
the development of future nanotube devices.

We therefore aim to propose the novel MR device 
by focusing on 
the following characteristic magnetic properties of the nanotube. 
First, 
a zigzag nanotube with a finite length appears to have 
spin-polarized edges, which are qualitatively same as 
those of a zigzag ribbon~\cite{Fujita}, 
because the zigzag nanotube 
corresponds just to the zigzag ribbon with short periodicity. 
It was theoretically shown~\cite{Fujita} that the zigzag ribbon 
forms spin polarized states with the ferrimagnetic order 
along zigzag edges, 
and the total magnetization of the ribbon is zero. 
Second, nanotubes encapsulating magnetic atoms 
such as Fe~\cite{particle1,particle2}, Co, or Ni atoms~\cite{FeCoNi} 
were recently fabricated.

In this paper, as the novel MR device with the nanotube, 
we propose 
``nonmagnetic metal (NM)/zigzag nanotube 
encapsulating magnetic atoms/NM'' junctions. 
Using the Green's function technique, 
we investigate the influence due to the spin-polarized edges 
and the encapsulated magnetic atoms on the transport. 
The MR effect due to the direction of spins of magnetic atoms 
appears reflecting the edges of the nanotube, 
and the magnitude of the MR ratio becomes much larger than 
the conventional ones in a certain parameter region.

\section{Model and Method}
Figure 1 shows a simplified model, 
in which the NM has a simple cubic structure, 
the $x$-direction of NMs is set to be semi-infinite, 
and their $yz$-directions have the periodic boundary condition 
meaning an infinite system. 
The number of unit cells in the circumference direction of the nanotube 
is 15, and the number of zigzag lines~\cite{Fujita} is 10. 
The each edge carbon atom of the nanotube is assumed to
interact with its nearest atom of the cubic lattice of the NM.

On the other hand, the encapsulated magnetic atoms have localized spins. 
We here assume that 
their spins are divided into several spin clusters 
which interact with their nearest carbon atoms respectively, 
and all the spin clusters have the identical total spin. 
The total spin is approximately represented by 
a classical spin ${\bf S}$ 
on the assumption that the spin cluster consists of many spins, 
and further every ${\bf S}$ is set to have the same direction. 
Now, each ${\bf S}$ is arrayed along 5-th and 6-th zigzag lines 
of the nanotube, and 
it has one to one interaction with carbon atoms in their lines. 
According to the theory on magnetic impurity problem~\cite{mag_imp}, 
we take into account antiferromagnetic exchange interactions 
between conduction electron spins and ${\bf S}$'s, 
where all the exchange interactions are set to be the same magnitude. 
\\
\\[-0.3cm]

\begin{figure}[ht]
\begin{center}
\resizebox{!}{3.5cm}{\includegraphics{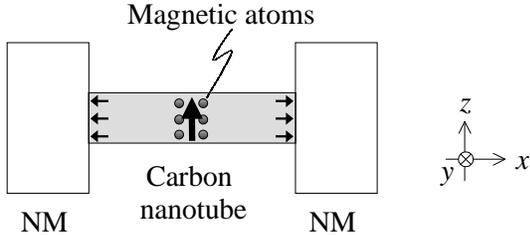}}\\
\vspace{0.3cm}
\caption{\footnotesize 
A schematic illustration of 
``NM/zigzag carbon nanotube encapsulating magnetic atoms/NM'' junctions, 
where electric currents flow in the $x$-direction. 
Arrows at the both edges of the nanotube represent spin-polarized states. 
}
\end{center}
\end{figure}

Based on a single orbital Hubbard model 
with nearest neighbor transfer integrals, 
the Hamiltonian is given by,
\begin{eqnarray}
\label{ham1} 
&& H=
   \sum_{i} \sum_\sigma e_{i,\sigma} c_{i,\sigma}^\dag c_{i,\sigma} 
\nonumber \\
&& \hspace{0.8cm} +\sum_{\langle i,j \rangle} \sum_\sigma 
     \left( t_{i,j} c_{i,\sigma}^\dag c_{j,\sigma} +{\rm h.c.} \right) 
\nonumber \\ 
&&\hspace*{0.8cm} + U \sum_{i \in {\rm CNT}} n_{i,\uparrow} n_{i,\downarrow} \nonumber \\
&&\hspace*{0.8cm} -J\sum_{i \in {\rm mag}}  \sum_{\sigma,\sigma'} 
{\bf \sigma}_{\sigma,\sigma'} \cdot {\bf S} 
c_{i,\sigma}^\dag c_{i,\sigma'}, 
\end{eqnarray}
where 
$c_{i,\sigma}$ ($c_{i,\sigma}^{\dag} $) is 
the annihilation (creation) operator of an electron 
with spin-$\sigma$ ($=\uparrow$ or $\downarrow$) at the $i$-th site, 
$n_{i,\sigma}$ is the corresponding number operator, 
$e_{i,\sigma}$ is the corresponding on-site energy, 
$t_{i,j}$ is the transfer integral 
between the $i$-th site and the $j$-th site, 
$U$ is the on-site Coulomb energy, 
and $J$ is the antiferromagnetic exchange integral 
with negative sign~\cite{mag_imp}. 
Furthermore, ${\bf \sigma}_{\sigma,\sigma'}$ is 
the $(\sigma,\sigma')$ component of the Pauli matrix 
for the conduction electron spin, 
and ${\bf S}$ $[=(S_{x},S_{y},S_{z})]$ represents 
the classical spin with $S \equiv |{\bf S}|$. 
Each ${\bf S}$ is considered to exist parallel to $xz$-plane, 
and an angle between ${\bf S}$ and $x$-axis 
is written as $\theta$. 
$\sum_{i \in {\rm CNT}}$ 
($\sum_{i \in {\rm mag}}$) means that the summation is taken 
for the nanotube (carbon atoms interacting with magnetic atoms).

We treat the term with on-site Coulomb energy 
within the mean field approximation~\cite{Fujita}. 
Furthermore, in this model, the magnetization of edges of the nanotube
is assumed to be pinned parallel to the $x$-axis, 
by applying an exchange bias 
to either the left edge or the right one of the nanotube. 
The exchange bias is supposed to arise from, 
for example, magnetic particles 
attached outside the rim of the nanotube edge.

Using the Green's function technique~\cite{cond,kokado3}, 
we calculate the conductance at zero temperature, 
\begin{eqnarray}
\Gamma (\theta,U)=\sum_{\sigma,\sigma'} 
\Gamma_{\sigma,\sigma'} (\theta,U), 
\end{eqnarray}
and
\begin{eqnarray}
&&\Gamma_{\sigma,\sigma'}(\theta,U)= \nonumber \\
&&\frac{4 \pi^2 e^2 }{h}
\displaystyle{\sum_{i,j,m,n}}
\langle i,\sigma| \hat{D}_L | j,\sigma \rangle 
\langle j,\sigma| \hat{T}^\dag (\theta,U)| m,\sigma' \rangle \nonumber \\
&&\hspace*{1cm}\times \langle m,\sigma'| \hat{D}_R | n,\sigma' \rangle 
\langle n,\sigma'| \hat{T} (\theta,U)| i,\sigma \rangle, 
\end{eqnarray}
where $i,j,m,n$ are coordinates of contact points in NMs, 
$\hat{D}_{L(R)}$ is the density-of-states (DOS) operator 
at the Fermi level $E_{\mbox{\tiny F}}$ of the left NM (right NM), 
and $\hat{T}(\theta,U)$ is the $T$-matrix with 
$G(\theta,U)=(E_{\mbox{\tiny F}} + {\rm i} 0 - H)^{-1}$
and couplings between NMs and the nanotube. 
Here, $\Gamma_{\sigma,\sigma'}(\theta,U)$ is the conductance 
for the transmission from the spin-$\sigma$ state of the left NM 
to the spin-$\sigma'$ one of the right NM. 
Furthermore, we calculate the MR ratio, 
defined by 
\begin{eqnarray}
R_{MR}(\theta,U)=100\times 
\frac{\Gamma (\theta,U)-\Gamma (\pi/2,U)}{\Gamma (\pi/2,U)}. 
\end{eqnarray}

In this calculation, 
we set $t_{i,j}=t~(<0)$~\cite{transfer} and $v_{i,j}=0.1t$, 
assuming that $v$ is smaller than $t$ 
because of different types between two orbitals, 
imperfect lattice matches at the interface, and so on. 
The on-site energy of the carbon atom 
$e_{i,\uparrow}/|t|$ (=$e_{i,\downarrow}/|t|$) is set to be 0 
by focusing on its $\pi$ orbital, 
and then 
$e_{i,\uparrow}/|t|$ (=$e_{i,\downarrow}/|t|$) of the both NMs is 5.4 
by considering that the $s$-orbital 
contributes to the transport of the both NMs. 
Furthermore, $E_{\mbox{\tiny F}}$ is self-consistently determined 
so as to keep half filling in the nanotube 
for the respective parameter sets of $U$ and $\theta$.

\section{Calculated Results and Considerations}
We first investigate magnetic properties 
of the carbon nanotube for the finite $U/|t|$ and $JS/|t|$=0. 
The zigzag nanotube with the finite length forms spin-polarized states 
with the ferrimagnetic order along zigzag edges, 
and the total magnetization of the ribbon is zero 
reflecting an opposite spin polarization between both edges. 
Such the feature has been seen in the ribbon~\cite{Fujita}, too. 
Also, 
when the spin polarization~\cite{Teresa} 
at the left (right) edge of the nanotube 
is defined by 
\begin{eqnarray}
P_{L(R)}= 
\frac{D_{L(R),\uparrow} - D_{L(R),\downarrow}}
{D_{L(R),\uparrow} + D_{L(R),\downarrow}}, 
\end{eqnarray}
where $D_{L(R),\uparrow}$ is the local DOS 
at the left (right) edge for the spin-$\uparrow$ at $E_{\mbox{\tiny F}}$ 
and $D_{L(R),\downarrow}$ is 
that for the spin-$\downarrow$ at $E_{\mbox{\tiny F}}$, 
we obtain $P_L >0$ and $P_R = -P_L$. 
\\
\\[-0.3cm]

\begin{figure}[ht]
\begin{center}
\resizebox{!}{8cm}{\includegraphics{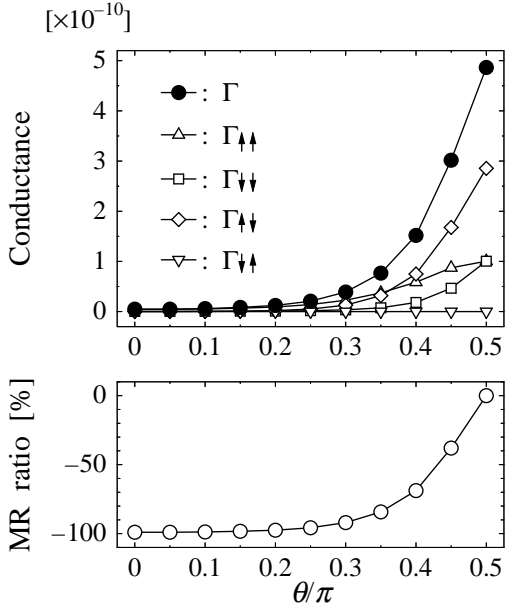}}\\
\vspace{0.3cm}
\caption{\footnotesize 
Upper panel: $\Gamma(\theta,|t|)$ and 
$\Gamma_{\sigma,\sigma'}(\theta,|t|)$ vs $\theta$. 
Lower panel: 
The MR ratio $R_{MR}(\theta,|t|)$ vs $\theta$. 
The unit of $4 \pi^2 e^2 /h$=1 is adopted. 
}
\end{center}
\end{figure} 
In the upper panel of Fig. 2, we show 
$\theta$ dependence of $\Gamma(\theta,U)$ and 
$\Gamma_{\sigma,\sigma'}(\theta,U)$ 
with $U/|t|$=1 and $JS/|t|$=$-$0.5. 
When $\theta$ is changed from 0 to $\pi/2$, 
$\Gamma(\theta,|t|)$ increases 
because of the enhancement of the spin-flip transport. 
Here, the spin-flip transport is understood 
by focusing on the main component 
based on the relations of $P_L >0$ and $P_R=-P_L$ 
as follows: 
First, the spin-$\uparrow$ of the electron from the left edge of the nanotube 
is flipped by spins of magnetic atoms with $\theta \ne 0$ 
in the center of the nanotube, 
and it changes into the spin-$\downarrow$. 
Second, the electron flows to the right edge with conservation of the spin. 
In fact, 
$\Gamma_{\uparrow,\downarrow}(\theta,|t|)$ rapidly increases with $\theta$, 
while $\Gamma_{\downarrow,\uparrow}(\theta,|t|)$ 
has little $\theta$ dependence 
and 
$\Gamma_{\uparrow,\uparrow}(\theta,|t|)$ and 
$\Gamma_{\downarrow,\downarrow}(\theta,|t|)$ 
moderately increase with $\theta$. 
It should be noted here that 
such $\theta$ dependences are regarded as the MR effect 
due to the direction of spins of magnetic atoms.

As seen from the lower panel of Fig. 2, 
$R_{MR}(\theta,U)$ with the same $U/|t|$ and $JS/|t|$ strongly depends on 
$\theta$ owing to the spin-flip transport. 
We emphasize that 
the large MR ratio of about 100\% can be found 
between $\theta$=0 and $\pi$/2. 
\\
\\[-0.3cm]

\begin{figure}[ht]
\begin{center}
\resizebox{!}{10.2cm}{\includegraphics{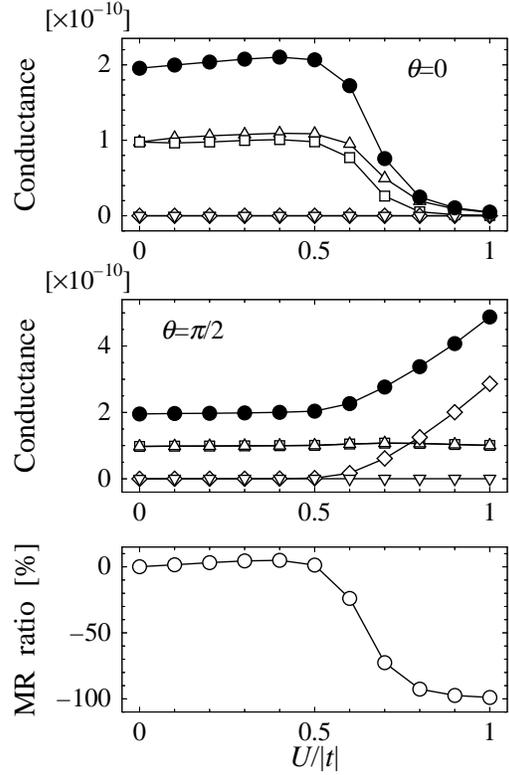}}\\
\vspace{0.3cm}
\caption{ \footnotesize 
Upper panel: $\Gamma(0,U)$ and 
$\Gamma_{\sigma,\sigma'}(0,U)$ vs $U/|t|$. 
Middle panel: $\Gamma(\pi/2,U)$ and 
$\Gamma_{\sigma,\sigma'}(\pi/2,U)$ vs $U/|t|$. 
The meanings of dots in upper and middle panels follow 
those of the upper panel of Fig. 2. 
Lower panel: The MR ratio $R_{MR}(0,U)$ vs $U/|t|$. 
The unit of $4 \pi^2 e^2 /h$=1 is adopted. 
}
\end{center}
\end{figure} 
In the upper (middle) panel of Fig. 3, we consider 
$U/|t|$ dependence of $\Gamma(\theta,U)$ and 
$\Gamma_{\sigma,\sigma'}(\theta,U)$ 
with $\theta$=0 ($\theta$=$\pi$/2) in the case of $JS/|t|$=$-$0.5. 
Difference between 
$\Gamma(0,0)$ and $\Gamma(\pi/2,0)$ is not present. 
For $U/|t| \le 0.5$, 
$\Gamma(0,U)$ and $\Gamma(\pi/2,U)$ 
have little dependence on $U/|t|$. 
When $U/|t|$ increases from 0.5, 
$\Gamma(\pi/2,U)$ increases and $\Gamma(0,U)$ decreases. 
The behavior indicates that 
$|P_{L}|$ and $|P_{R}|$ increase 
with increasing $U/|t|$ for $U/|t| >0.5$ 
and therefore the spin-flip transport becomes dominant. 
Note that 
for $U/|t| >0.5$, 
$\Gamma_{\uparrow,\downarrow}(\pi/2,U)$ increases with $U/|t|$, 
while $\Gamma_{\uparrow,\uparrow}(0,U)$ and 
$\Gamma_{\downarrow,\downarrow}(0,U)$ decrease with $U/|t|$. 
Also, 
the magnitude of $R_{MR}(\theta,U)$ with $\theta$=0 
increases with $U/|t|$ for $U/|t| > 0.5$ 
and becomes about 100\% for $U/|t| > 0.8$, 
although it is very small for $U/|t| \le 0.5$ 
(see the lower panel of Fig. 3).

\section{Comments}
Comparing with conventional FM/carbon nanotube/FM 
junctions~\cite{Tsukagoshi,Zhao,Zhao2,Kim,Mehrez}, 
we have found that in a certain parameter region, 
the MR ratio 
due to the direction of spins of the encapsulated magnetic atoms 
of the present junctions becomes larger than MR ratios~\cite{MR} 
of 9\%~\cite{Tsukagoshi}, 23\%~\cite{Zhao}, and 26\%~\cite{Zhao2} 
of the conventional ones. 
However, 
more detailed studies including evaluation of parameters 
should be necessary in future.

As realistic junctions relevant to the present model, 
we point out, for example, 
``Ti/zigzag carbon nanotube encapsulating Fe atoms/Ti" junctions, 
where zigzag edges of the nanotube are 
successfully contacted with Ti electrodes. 
In fact, in the recent experiments, 
Ti/carbon nanotube/Ti junctions were 
applied to the field-effect transistor~\cite{Ti},  
and also carbon nanotubes encapsulating 
Fe~\cite{particle1,particle2}, Co, or Ni atoms~\cite{FeCoNi} 
were often fabricated.

\section{Conclusion}
As the novel MR device with the carbon nanotube, 
we proposed 
``NM/zigzag carbon nanotube encapsulating magnetic atoms/NM'' junctions. 
By theoretically investigating the spin-dependent transport, 
we found that 
the MR effect due to the direction of spins of magnetic atoms 
appears reflecting spin-polarized edges of the nanotube. 
In the case of $U/|t|> 0.8$, 
the magnitude of the MR ratio becomes about 100\%, 
which is much larger than those of the conventional junctions. 
We anticipate that such junctions will be applied to 
magnetic field sensors 
with high sensitivity. 
\\
\\[-0.1cm]

\newpage
\hspace*{-0.3cm}{\bf Acknowledgements}\\
\\
This work has been supported by 
Special Coordination Funds for Promoting Science and Technology, Japan. 
One of the authors (K.H.) acknowledges 
the partial financial support by NEDO 
under the Nanotechnology Program, too.


\begin{thebibliography}{9}
\bibitem{MR}
The MR effect of FM/nanotube/FM junctions corresponds to 
the difference of conductances between parallel (P) magnetization 
and anti-parallel (AP) one of FMs. 
Its MR ratio is defined by 
$100 \times (\Gamma_P - \Gamma_{AP})/\Gamma_{P}$ 
with $\Gamma_{P(AP)}$ being 
the conductance of the P (AP) case.
\bibitem{transfer}
The transfer integral $t$ was set to be $-$2.75 eV in ref.\cite{Mehrez}.
\end{thebibliography}
\end{document}